\documentclass[twocolumn,english,aps]{revtex4-2}
\usepackage[utf8]{inputenc}
\usepackage{float}
\usepackage{amsmath}
\usepackage{amssymb}
\usepackage{graphicx}

\makeatletter

\DeclareTextSymbolDefault{\textquotedbl}{T1}

\usepackage[margin=15mm]{geometry}
\usepackage{babel}
\usepackage{bbold}

\usepackage{hyperref}
\hypersetup{
	colorlinks = true,
	linkcolor=[rgb]{0.717647, 0.0588235, 0.227451},
	citecolor=[rgb]{0.717647, 0.0588235, 0.227451},
	urlcolor=[rgb]{0.364706, 0.501961, 0.705882}
}

\makeatother

\begin{document}
%\title{Theory of birefringence due to magnon-phonon coupling in a bilayer structure}
\title{Theory of polarization-dependent phonon pumping in ferromagnetic/non-magnetic bilayers}
\author{M. Cherkasskii}
\email{macherkasskii@hotmail.com}

\affiliation{Institute for Theoretical Solid State Physics, RWTH Aachen University, DE-52074 Aachen, Germany}
\author{F. Engelhardt}
\affiliation{Institute for Theoretical Solid State Physics, RWTH Aachen University, DE-52074 Aachen, Germany}
\affiliation{Max Planck Institute for the Science of Light, 91058 Erlangen, Germany}
\affiliation{Department of Physics, University Erlangen-Nuremberg, 91058 Erlangen, Germany}
\author{M. M{ü}ller}
\affiliation{Walther-Mei{ß}ner-Institut, Bayerische Akademie der Wissenschaften, 85748 Garching, Germany}
\affiliation{Technical University of Munich, TUM School of Natural Sciences, Physics Department, 85748 Garching, Germany}
\author{J. Weber}
\affiliation{Walther-Mei{ß}ner-Institut, Bayerische Akademie der Wissenschaften, 85748 Garching, Germany}
\affiliation{Technical University of Munich, TUM School of Natural Sciences, Physics Department, 85748 Garching, Germany}
\author{S. T. B. Goennenwein}
\affiliation{Department of Physics, University of Konstanz, 78457 Konstanz, Germany}

\author{M. Althammer}
\affiliation{Walther-Mei{ß}ner-Institut, Bayerische Akademie der Wissenschaften, 85748 Garching, Germany}
\affiliation{Technical University of Munich, TUM School of Natural Sciences, Physics Department, 85748 Garching, Germany}
\author{H. Huebl}
\affiliation{Walther-Mei{ß}ner-Institut, Bayerische Akademie der Wissenschaften, 85748 Garching, Germany}
\affiliation{Technical University of Munich, TUM School of Natural Sciences, Physics Department, 85748 Garching, Germany}
\affiliation{Munich Center for Quantum Science and Technology (MCQST), 80799 Munich, Germany}
\author{S. {Viola Kusminskiy}}
\affiliation{Institute for Theoretical Solid State Physics, RWTH Aachen University, DE-52074 Aachen, Germany}
\affiliation{Max Planck Institute for the Science of Light, 91058 Erlangen, Germany}
\begin{abstract}
We develop a theoretical model for polarization-selective phonon pumping induced by magnon-phonon coupling in a ferromagnetic/non-magnetic acoustic bilayer structure, focusing on the effects arising from a misalignment between the magnetic and crystallographic symmetry axes. Our model considers the coupled equations of motion describing uniform magnetization dynamics (the Kittel mode) and elastic waves in both layers, incorporating phonon pumping and boundary conditions at the interface. We show that even small misalignments lift the degeneracy of transverse shear elastic modes, resulting in phononic birefringence characterized by distinct propagation velocities for linearly polarized modes. Furthermore, our analysis reveals that magnon-phonon hybridization gives magnetic-field-dependent properties to otherwise non-magnetic phonons. We show that the polarization transfer between linearly polarized phonons and the circularly polarized Kittel mode can be tuned with an external magnetic field. Our theoretical results quantitatively reproduce recent experimental findings~\citep{Muller2024Chiral}. 
\end{abstract}
\maketitle

\section{Introduction}

It was shown by C. Kittel~\citep{kittel_interaction_1958} that the coupling between magnons and phonons can be significant when their frequencies and wavelengths are matched. His work set the stage for a plethora of studies on the properties and possible applications of magnetoelastic effects~\citep{Lacheisserie1993}. Several studies have focused on the coupling between elastic and spin waves at non-zero wave vectors \citep{schlomann_generation_2004,schlomann_generation_2004-1,rezende_magnetoelastic_2003}. In this context, the boundaries of magnetoelastic samples were considered as the origin of standing magnetoelastic waves, also known as ferromagnetoelastic resonance \citep{kobayashi1973ferromagnetoelastic}. However, the discovery of phonon pumping changed the perspective \citep{Streib2018}. It was demonstrated that when a ferromagnetic film is attached to an elastic nonmagnetic material, the magnetization dynamics changes, acquiring a dependence on the geometrical parameters of the heterostructure.  As the magnetization precesses, it pumps phonons into the nonmagnetic layer, enabling an effective coupling of the Kittel mode in the magnetic material with elastic waves throughout the structure. In particular, a new dissipative channel opens for the magnetization due to the pumped phonons. Phonon pumping has been further explored in the context of angular momentum transfer within heterostructures \citep{an_coherent_2020,an_bright_2022,sato_dynamic_2021,schlitz_magnetization_2022}. Interference between the ferromagnetic resonances of two yttrium iron garnet samples was observed over a macroscopic distance, mediated by the exchange of coherent shear waves propagating through a nonmagnetic gadolinium gallium garnet slab \citep{an_coherent_2020}. Coherent coupling between two macrospins via phonons in this kind of structures was observed, which is crucial for designing quantum devices
harnessing angular momentum transfer \citep{an_bright_2022}. Additionally, it was theoretically shown that phonon pumping can be utilized to transfer incoherent spins \citep{Ruckriegel2020}. For practical applications, the magnetoelastic coupling itself has been recently proposed as a means to design quantum magnetic memory cells \citep{chu_creation_2018,hann_hardware-efficient_2019,wallucks_quantum_2020}, sensors \citep{bienfait_quantum_2020,satzinger_quantum_2018,pottsMagnonPhononQuantumCorrelation2020a, potts_dynamical_2021}, and transducers \citep{han_superconducting_2022,jiang_efficient_2020,arnold_converting_2020,engelhardt_optimal_2022}. These highlight the significant interest in phonon-magnon processes and their potential in various technological applications.

A recent study~\citep{Muller2024Chiral, MullerTemperature2024} using low-acoustic-damping
%materials sapphire and silicon crystalline 
substrates interfaced with ferromagnetic thin films at low temperatures provided insights into the effect of crystal symmetry on phonon pumping. 
%In particular, the excitation of chiral phonons was studied.
The experimental observation of transverse shear waves with differing propagation velocities suggests that settings enabling the investigation of phononic birefringence can be realized. Using a minimal theoretical model it was possible to capture the main experimental features, however the model did not fully incorporate phonon pumping. Whereas T. Sato \emph{et al.}~\citep{sato_dynamic_2021, schlitz_magnetization_2022} explored the role of phonon pumping and boundary conditions for bilayers in detail, their model assumed an alignment between magnetic and elastic symmetry axes, leading to a degeneracy in the transverse shear modes. To bridge this gap, in this work we develop a theory that  accounts for phonon pumping in magnetic/nonmagnetic bilayers in the presence of a misalignment between magnetic and elastic symmetry axes. We benchmark our model against the experimental results presented in Ref.~\citep{Muller2024Chiral}, showing that it quantitatively describes the magnon-phonon hybridization features as measured in microwave transmission. Our model serves as a basis to treat the magnetoelastic dynamics in bilayers with different crystalline symmetries and presenting misalignments, and therefore represents an important step for the understanding of phononic birefringence using this excitation scheme. These effects are of particular importance to understand and tailor the transfer and transport of angular momentum via phonons in these structures. To this end, we analyze the polarization transfer between linearly polarized phonons and the circularly polarized Kittel mode and show that it can be controlled by an external magnetic field.

\begin{figure}
\begin{centering}
\includegraphics{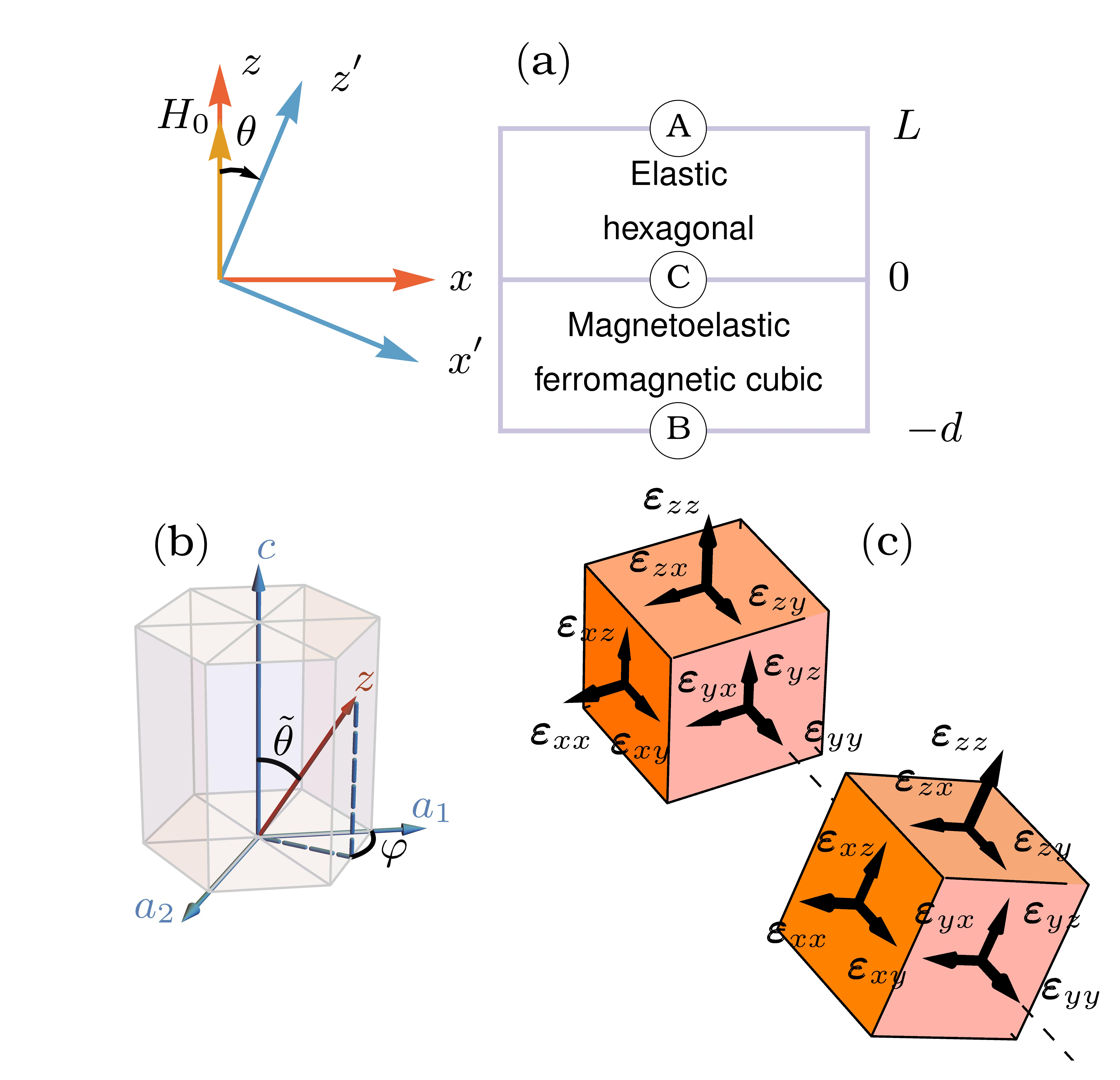} 
\par\end{centering}
\caption{(a) The orientation of the coordinate systems and schematic representation of the two-layered structure composed of a ferromagnetic film with thickness $d$ and an elastic insulator with thickness $L\gg d$. The boundaries and interfaces are labeled by letters A, B and C. (b) Propagation direction of the elastic waves along $z$-axes with respect to the crystallgraphic axes of the hexagonal layer, $\theta=-\tilde{\theta}$. (c) Rotation of strain tensor.}
\label{fig:str_rotation} 
\end{figure}

The manuscript is structured as follows. In Sec.~\ref{sec:EoM} we discuss the coupled equations of motion for a bilayer structure as shown in Fig.~\ref{fig:str_rotation}(a), starting from the magnetoelastic energy density. In Sec.~\ref{sec:Velocity-split} we demonstrate how the rotation of the nonmagnetic layer with respect to the magnetic crystalline axes leads to non-degenerate linearly polarized phonon modes, i.e. birefringence, and we calculate the dispersion of the ensuing elastic waves. In Sec.~\ref{sec:Power-abs} we calculate the power absorption in terms of the magnetic susceptibility and discuss its comparison with the experimental results of Ref.~\citep{Muller2024Chiral}. In particular, we focus on the effect of phonon pumping and the boundary conditions on the susceptibility. In Sec.~\ref{sec:Polarization} we study the polarization transfer between the elastic waves and the Kittel mode. In Sec.~\ref{sec:conclusions} we present the conclusions and an outlook. We relegate lenghty calculations to App.~\ref{sec:appendix-Velocity-split} and provide further discussion on the effects of different lattice symmetries in App.~\ref{sec:hex-cubi}.

\section{Equations of motion}

\label{sec:EoM}
In this section, we set up the equations of motion describing the coupled dynamics of magnetic and elastic excitations in a bilayer structure consisting of a nonmagnetic substrate and a ferromagnetic film as depicted in {[}Fig.~\ref{fig:str_rotation}(a){]}. The former has elastic properties, whereas the latter is magnetoelastic. Both components support the propagation of elastic waves. We consider standing elastic waves along the entire structure's thickness and uniform magnetization dynamics (ferromagnetic resonance, FMR) in the magnetic film. Hence, the dynamical variables are the strain field ${\bf u}$ and the magnetization ${\bf M}$. The dynamics of the system is governed by two coupled equations of motion. Hooke's law 
\begin{equation}
\rho\partial_{tt}{\bf u}+\eta\partial_{t}{\bf u}={\bf f}^{\text{tot}},\label{eq:elastic_gen}
\end{equation}
governs the dynamics of the strain fields, where $\rho$ is the mass density, $\eta$ is the elastic damping, and ${\bf f}^{\text{tot}}$ is the total force per unit volume, which includes both elastic and magnetic contributions. The magnetization dynamics is described by the Landau--Lifshitz--Gilbert (LLG) equation
\begin{equation}
\partial_{t}{\bf M}=-\gamma\mu_{0}{\bf M}\times{\bf H}^{{\rm eff}}+\frac{\alpha}{M_{{\rm s}}}{\bf M}\times\partial_{t}{\bf M},\label{eq:LLG_gen}
\end{equation}
where $\gamma$ is the absolute value of gyromagnetic ratio~\footnote{Note that we take $\gamma>0$ for definiteness. Eq. \eqref{eq:LLG_gen} with $\gamma>0$ implies a negative gyromagnetic ratio. The results can be easily generalized for an arbitrary sign of $\gamma$.}, $\alpha$ denotes the Gilbert damping, $\mu_{0}$ is the vacuum permeability, and $M_{{\rm s}}$ is the saturation magnetization. ${\bf H}^{{\rm eff}}$ is the effective magnetic field, with magnetic and magnetoelastic contributions. In what follows we will derive ${\bf f}^{\text{tot}}$ and ${\bf H}^{{\rm eff}}$ taking into account that the magnetic, elastic and magnetoelastic phenomena are anisotropic with specific symmetry axes. 

The magnetization precesses around the effective magnetic field, which includes the external applied magnetic field ${\bf H}_{0}$. In this work, we set  ${\bf H}_{0}$ to be orthogonal to the plane of the bilayer structure, see {[}Fig.~\ref{fig:str_rotation}(a){]}. In the linear regime, one considers small angles of precession, therefore the magnetization is almost aligned along ${\bf H}^{\rm eff}$. Fundamentally, elastic waves are influenced by the crystal symmetry, which can lead to anisotropic velocities~\citep{nye1985physical}. 
The magnetoelastic dynamics are determined by the symmetry of the magnetoelastic coupling energy, defined by the parallel and perpendicular magnetoelastic constants.

We consider the general case in which the symmetry axes of magnetic and elastic layers are not neccessarily aligned, which is of particular interest for the investigation of phononic birefringence. In an experiment, the symmetry axes are often slightly misaligned e.g. due imperfections in the crystal cutting and polishing process, i.e. the lattice vectors have a finite angle with respect to the sample geometry/sample surface. For specificity and to compare with the experimental results of  Ref.~\citep{Muller2024Chiral}, we assume that the ferromagnetic film is described by a strain tensor with cubic symmetry, whereas the substrate has hexagonal symmetry. The case of hexagonal ferromagnets is discussed in the Appendix~\ref{sec:hex-cubi}. The coordinate frame $\{x',y',z'\}$ for the strain tensor in the elastic layer is rotated by a small angle $\theta$ around the \emph{y-}axis to account for the misalignment of magnetic and elastic symmetries. This rotation results in the splitting of the degeneracy of the two linearly polarized transverse elastic waves.   The model can be directly generalized by including the rotation of the strain tensor of other heterostructure layers. However, the experimental results are adequately described by considering only the rotation of the elastic layer's tensor, as the rotation of the ferromagnetic layer's tensor has a negligible effect in the studied heterostructure. To elucidate the coupling between Eqs.~(\ref{eq:elastic_gen}) and (\ref{eq:LLG_gen}) and to model the effect of the axes misalignment we introduce the strain and stress tensors below.

\subsection{Elastic and magnetoelastic dynamics}

Due to the elastic and magnetoelastic effects, the total force density is the sum of two terms~\citep{Lacheisserie1993}
\begin{equation}
f_{i}^{\text{tot}}=f_{i}^{\text{el}}+f_{i}^{\text{mel}}=\sum_{j}\frac{\partial\sigma_{i,j}^{{\rm el}}}{\partial j}+\sum_{j}\frac{\partial\sigma_{i,j}^{{\rm mel}}}{\partial j}.\label{eq:f_tot}
\end{equation}
The variables $i$ and $j$ represent the three coordinates $x,$ $y,$ and $z$. In addition, we introduce the elastic $\sigma^{{\rm el}}$ and magnetoelastic $\sigma^{{\rm mel}}$ stress tensors. The gradient of the stress tensor quantifies the internal forces that neighboring points of a continuum solid exert on each other.

First, we consider the elastic stress, and note that the explicit form of the corresponding tensor depends on the crystal symmetry. Whereas here we focus on a crystalline substrate with hexagonal symmetry and a magnet with cubic symmetry, the method can be extended to other lattice symmetries. In general, the stress tensor is related to the strain tensor $\varepsilon$ according to the equation~\citep{Lacheisserie1993}  
\begin{equation}
\sigma_{i,j}^{{\rm el}}=\sum_{kl}c_{i,j,k,l}\varepsilon_{k,l},\label{eq:sigma_eps_gen}
\end{equation}
where $c_{i,j,k,l}$ are the components of the fourth-order tensor of stiffness. Owing to symmetry, many of the tensor components are either identical to each other, or equal to zero. It is common to employ the Voigt notation, and write Eq.~(\ref{eq:sigma_eps_gen}) for an hexagonal crystal as 
\begin{equation}
\begin{aligned} & \sigma_{x,x}^{{\rm el}}=c_{1,1}\varepsilon_{x,x}+c_{1,2}\varepsilon_{y,y}+2c_{1,4}\varepsilon_{y,z}+c_{1,3}\varepsilon_{z,z},\\
 & \sigma_{y,y}^{{\rm el}}=c_{1,2}\varepsilon_{x,x}+c_{1,1}\varepsilon_{y,y}-2c_{1,4}\varepsilon_{y,z}+c_{1,3}\varepsilon_{z,z},\\
 & \sigma_{z,z}^{{\rm el}}=c_{1,3}\varepsilon_{x,x}+c_{1,3}\varepsilon_{y,y}+c_{3,3}\varepsilon_{z,z},\\
 & \sigma_{y,z}^{{\rm el}}=\sigma_{z,y}^{{\rm el}}=c_{1,4}\varepsilon_{x,x}-c_{1,4}\varepsilon_{y,y}+2c_{4,4}\varepsilon_{y,z},\\
 & \sigma_{x,z}^{{\rm el}}=\sigma_{z,x}^{{\rm el}}=2c_{1,4}\varepsilon_{x,y}+2c_{4,4}\varepsilon_{x,z},\\
 & \sigma_{x,y}^{{\rm el}}=\sigma_{y,x}^{{\rm el}}=\left(c_{1,1}-c_{1,2}\right)\varepsilon_{x,y}+2c_{1,4}\varepsilon_{x,z},
\end{aligned}
\label{eq:sigma_el_hex_eps}
\end{equation}
i.e. we encode four indices $i,j,k,l$ in a short form \citep{nye1985physical}. To be precise, we consider the rhombohedral (I) class (Laue class $\overline{3}m$) and simplify it to the hexagonal class using $2c_{4,4}=c_{1,1}-c_{1,2}$ for the substrate. For the cubic ferromagnetic crystal it suffices to replace the coefficients 
\begin{equation}
c_{1,4}\rightarrow0,\;c_{3,3}\to c_{1,1},\;c_{1,3}\to c_{1,2}\label{eq:fromHexToCubic}
\end{equation}
in Eqs.~(\ref{eq:sigma_el_hex_eps}) to obtain
\begin{equation}
\begin{aligned} & \sigma_{x,x}^{{\rm el}}=c_{1,1}\varepsilon_{x,x}+c_{1,2}\varepsilon_{y,y}+c_{1,2}\varepsilon_{z,z},\\
 & \sigma_{y,y}^{{\rm el}}=c_{1,2}\varepsilon_{x,x}+c_{1,1}\varepsilon_{y,y}+c_{1,2}\varepsilon_{z,z},\\
 & \sigma_{z,z}^{{\rm el}}=c_{1,2}\varepsilon_{x,x}+c_{1,2}\varepsilon_{y,y}+c_{1,1}\varepsilon_{z,z},\\
 & \sigma_{y,z}^{{\rm el}}=\sigma_{z,y}^{{\rm el}}=2c_{4,4}\varepsilon_{y,z},\\
 & \sigma_{x,z}^{{\rm el}}=\sigma_{z,x}^{{\rm el}}=2c_{4,4}\varepsilon_{x,z},\\
 & \sigma_{x,y}^{{\rm el}}=\sigma_{y,x}^{{\rm el}}=2c_{4,4}\varepsilon_{x,y}.
\end{aligned}
\label{eq:sigma_el_cubic_eps}
\end{equation}
The stiffness coefficients are assumed to be given from experimental data. In the following we will restrict our model to the linear, symmetric strain tensor given by 
\begin{equation}
\varepsilon_{i,j}=\frac{1}{2}\left(\frac{\partial u_{j}}{\partial i}+\frac{\partial u_{i}}{\partial j}\right).\label{eq:eps_to_u}
\end{equation}

Second, to obtain the magnetoelatic stress tensor we consider the magnetoelastic energy density in the cubic ferromagnetic crystal
\begin{alignat}{1}
U^{\text{mel}} & =\frac{B_{\perp}}{M_{{\rm s}}^{2}}\left(M_{x}M_{y}\varepsilon_{x,y}+M_{y}M_{x}\varepsilon_{y,x}+M_{x}M_{z}\varepsilon_{x,z}\right.\nonumber \\
 & \left.+M_{z}M_{x}\varepsilon_{z,x}+M_{y}M_{z}\varepsilon_{y,z}+M_{z}M_{y}\varepsilon_{z,y}\right)\label{eq:Umel}\\
 & +\frac{B_{\parallel}}{M_{{\rm s}}^{2}}\left(\varepsilon_{x,x}M_{x}{}^{2}+\varepsilon_{y,y}M_{y}{}^{2}+\varepsilon_{z,z}M_{z}{}^{2}\right),\nonumber 
\end{alignat}
where $B_{\perp}$ and $B_{\parallel}$ are the magnetoelastic constants ~\citep{Abrahams1952, KittelRevModPhys1953, Kittel1958}. The magnetoelastic stress tensor is defined as
\begin{equation}
\sigma_{i,j}^{{\rm mel}}=\frac{\partial U^{\text{mel}}}{\partial\varepsilon_{i,j}}.\label{eq:sigma_mel_Umel}
\end{equation}

Note that we disregard rotational deformations and thus the additional magnetoelastic coupling that arises due to the magnetocrystalline anisotropy. We assume that this contribution to the magnetoelastic dynamics is negligibly small. This approximation is valid if the magnetoelastic constants $B_{\perp}$ and $B_{\parallel}$ are  sufficiently smaller than the uniaxial anisotropy~\citep{KittelRevModPhys1953, Kittel1958, knupfer2011domain}.

\subsection{Rotation of the strain tensor}

It is generally assumed that the symmetry axes of elastic and magnetic phenomena coincide. However, as previously noted, such alignment was not observed experimentally~\citep{Muller2024Chiral}. To reconcile theory with experimental data, we introduce the auxiliary Cartesian coordinate system $\left\{ x',y',z'\right\} $, and rotate the strain tensor {[}Fig.~\ref{fig:str_rotation}(c){]} 
\begin{equation}
\varepsilon'=R\varepsilon R^{T},\label{eq:epsRotation}
\end{equation}
where the Euler matrix is given by
\begin{equation}
R=\left(\begin{array}{ccc}
\cos\theta & 0 & \sin\theta\\
0 & 1 & 0\\
-\sin\theta & 0 & \cos\theta
\end{array}\right).\label{eq:R}
\end{equation}
This matrix describes the rotation of the \emph{z} and \emph{x}-axes by $\theta$, see {[}Fig.~\ref{fig:str_rotation}(b){]}. Note that we substitute $\varepsilon$ with $\varepsilon'$ exclusively in the elastic Eqs.~(\ref{eq:sigma_el_hex_eps},\ref{eq:sigma_el_cubic_eps}), while the magnetic component remains non-rotated. This rotation enables us to capture the effect of phononic birefringence.

\subsection{Magnetic dynamics}

Let us proceed to the description of magnetic phenomena. % We recall that magnetization is defined as the sum of magnetic moments divided by the volume of the solid. Conversely, the effective magnetic field is the field experienced by a given magnetic moment. The LLG Eq.~\ref{eq:LLG_gen} shows that the magnetization precessing around the direction of the effective magnetic field. This field, in turn, is related to the magnetic energy of the medium
The effective magnetic field is related to the magnetic energy density of the medium 
\begin{equation}
{\bf H}^{\text{eff}}=-\dfrac{1}{\mu_{0}}\dfrac{\partial U}{\partial{\bf M}},\label{eq:Heff_gen}
\end{equation}
which comprises several contributions~\citep{gurevich_magnetization_2020}. The Zeeman term is 
\begin{equation}
U^{{\rm Z}}=-\mu_{0}\left(H_{x}M_{x}+H_{y}M_{y}+H_{z}M_{z}\right),\label{eq:Uz}
\end{equation}
where $H_{x}$ and $H_{y}$ are small-signal alternating magnetic field, and $H_{z}=H_{0}$ is the constant magnetic field. The demagnetization energy density is given by
\begin{equation}
U^{\text{dm}}=\dfrac{1}{2}\mu_{0}\left(N_{xx}M_{x}{}^{2}+N_{yy}M_{y}{}^{2}+N_{zz}M_{z}{}^{2}\right),
\end{equation}
where $N_{xx},$ $N_{yy}$, and $N_{zz}$ are the demagnetization factors. Additionally we include a magnetocrystalline anisotropy term
\begin{equation}
U^{\text{uni}}=\frac{K_{1}}{M_{{\rm s}}^{2}}\left(M_{x}^{2}+M_{y}^{2}\right),\label{eq:Uuni}
\end{equation}
which we assume to be uniaxial for simplicity, where $\ensuremath{K_{1}}$ is the anisotropy constant. Note that, as per our assumption, the coordinate system in which the magnetic energy is defined coincides with the principal coordinate system of the film. In this system, the \emph{z}-axis is oriented orthogonally to the film surfaces. Finally, we add the magnetoelastic energy Eq.~(\ref{eq:Umel}). Thus, we can write the effective magnetic field explicitly as
\begin{widetext}
\begin{equation}
\begin{aligned}H_{x}= & -\frac{1}{\mu_{0}M_{{\rm s}}^{2}}\left[-\mu_{0}M_{{\rm s}}^{2}H_{x}+2K_{1}M_{x}+\mu_{0}M_{{\rm s}}^{2}N_{xx}M_{x}\right.\\
 & \left.+2B_{\|}M_{x}u_{x}^{(1,0,0)}+B_{\perp}\left(M_{z}u_{x}^{(0,0,1)}+M_{y}u_{x}^{(0,1,0)}+M_{y}u_{y}^{(1,0,0)}+M_{z}u_{z}^{(1,0,0)}\right)\right]\\
H_{y}= & -\frac{1}{\mu_{0}M_{{\rm s}}^{2}}\left[-\mu_{0}M_{{\rm s}}^{2}H_{y}+2K_{1}M_{y}+\mu_{0}M_{{\rm s}}^{2}N_{yy}M_{y}\right.\\
 & \left.+2B_{\|}M_{y}u_{y}^{(0,1,0)}+B_{\perp}\left(M_{z}u_{y}^{(0,0,1)}+M_{x}u_{x}^{(0,1,0)}+M_{z}u_{z}^{(0,1,0)}+M_{x}u_{y}^{(1,0,0)}\right)\right]\\
H_{z}= & -\frac{1}{\mu_{0}M_{{\rm s}}^{2}}\left[-\mu_{{\rm s}}M_{{\rm s}}^{2}H_{z}+\mu_{{\rm s}}M_{{\rm s}}^{2}N_{zz}M_{z}\right.\\
 & \left.+2B_{\|}M_{z}u_{z}^{(0,0,1)}+B_{\perp}\left(M_{x}u_{x}^{(0,0,1)}+M_{y}u_{y}^{(0,0,1)}+M_{y}u_{z}^{(0,1,0)}+M_{x}u_{z}^{(1,0,0)}\right)\right].
\end{aligned}
\label{eq:Heff}
\end{equation}
\end{widetext}
where we denoted spatial derivatives with superscripts, e.g. \begin{equation}
u_{z}{}^{(1,1,0)}=\dfrac{\partial^{2}u_{z}}{\partial x\partial y}\,\,\,, \,\,\,u_{z}{}^{(0,0,2)}=\dfrac{\partial^{2}u_{z}}{\partial z^{2}}\,... \,.
\end{equation}
Note that the exchange energy is not included in the effective field, since we are interested solely in the dynamics of the Kittel mode. 

\subsection{System of coupled motion equations}

We substitute Eqs.~(\ref{eq:f_tot}-\ref{eq:Heff}) into Eqs.~(\ref{eq:elastic_gen}) and (\ref{eq:LLG_gen}) to obtain a closed system of differential equations with respect to ${\bf u}$ and ${\bf M}$. In this work we focus on the standing elastic waves along the structure thickness and the uniform Kittel mode (specifically precessing along the z-axis in Fig.~\ref{fig:str_rotation}). Therefore the in-plane wave numbers can be set to zero $k_{x}=k_{y}=0,$ whereas the time dependence is given by $e^{{\rm i}\omega t}.$ Furthermore, we linearize the equations of motion by setting $M_z \approx M_{\mathrm{s}}$, implying small precession angles and a magnetization nearly parallel to the static part of $\mathbf{H}^{\mathrm{eff}}$. We then focus on the ferromagnetic mode dominated by the Zeeman field, treating the magnetization as aligned along the $z$-axis. The non-aligned case can be addressed analogously. We neglect the longitudinal mode since its contribution to phonon pumping is of second order in terms of the displacements and fluctuations of the magnetization. Thus, the equations for the non-magnetic crystal are given by
\begin{widetext}
\begin{alignat}{1}
 & \rho\omega^{2}u_{x}-{\rm i}\eta\omega u_{x}+c_{4,4}\cos\left(2\theta\right)\partial_{zz}u_{x}-c_{1,4}\sin\left(\theta\right)\partial_{zz}u_{y}-c_{4,4}\sin\left(2\theta\right)\partial_{zz}u_{z}=0,\label{eq:Hooke_LL_el1}\\
 & \rho\omega^{2}u_{y}-{\rm i}\eta\omega u_{y}-c_{1,4}\sin\left(\theta\right)\cos\left(\theta\right)\partial_{zz}u_{x}+c_{4,4}\cos\left(\theta\right)\partial_{zz}u_{y}+c_{1,4}\sin^{2}\left(\theta\right)\partial_{zz}u_{z}=0,\nonumber \\
 & \rho\omega^{2}u_{z}-{\rm i}\eta\omega u_{z}+\left(c_{3,3}-c_{1,3}\right)\sin\left(\theta\right)\cos\left(\theta\right)\partial_{zz}u_{x}+\left[c_{3,3}\cos^{2}\left(\theta\right)+c_{1,3}\sin^{2}\left(\theta\right)\right]\partial_{zz}u_{z}=0,\label{eq:Hooke_LL_el3}
\end{alignat}
and for the ferromagnet by
\begin{alignat}{1}
 & \tilde{\rho}\omega^{2}u_{x}-{\rm i}\tilde{\eta}\omega u_{x}+\tilde{c}_{4,4}\cos\left(2\theta\right)\partial_{zz}u_{x}-\tilde{c}_{4,4}\sin\left(2\theta\right)\partial_{zz}u_{z}=0\label{eq:Hooke_LL_mel1}\\
 & \tilde{\rho}\omega^{2}u_{y}-{\rm i}\tilde{\eta}\omega u_{y}+\tilde{c}_{4,4}\cos\left(\theta\right)\partial_{zz}u_{y}=0,\nonumber \\
 & \tilde{\rho}\omega^{2}u_{z}-{\rm i}\tilde{\eta}\omega u_{z}+\tilde{c}_{4,4}\sin\left(2\theta\right)\partial_{zz}u_{x}+\left[\tilde{c}_{1,1}-2\tilde{c}_{4,4}\sin^{2}\left(\theta\right)\right]\partial_{zz}u_{z}=0\label{eq:Hooke_LL_mel3}\\
 & -\gamma\mu_{0}H_{y}+{\rm i}\omega\dfrac{M_{x}}{M_{{\rm s}}}+\left(\dfrac{{\rm i}\alpha\omega}{M_{{\rm s}}}+\gamma\dfrac{2K_{1}}{M_{{\rm s}}^2}+\dfrac{\gamma\mu_{0}H_{0}}{M_{{\rm s}}}+\gamma\mu_{0}N_{yy}-\gamma\mu_{0}N_{zz}\right)M_{y}+\dfrac{B_{\perp}}{M_{{\rm s}}}\gamma\partial_{z}u_{y}=0,\label{eq:LL_lin_Mx}\\
 & \gamma\mu_{0}H_{x}-\left(\dfrac{{\rm i}\alpha\omega}{M_{{\rm s}}}+\gamma\dfrac{2K_{1}}{M_{{\rm s}}^2}+\dfrac{\gamma\mu_{0}H_{0}}{M_{{\rm s}}}+\gamma\mu_{0}N_{xx}-\gamma\mu_{0}N_{zz}\right)M_{x}+\dfrac{{\rm i}\omega M_{y}}{M_{{\rm s}}}-\dfrac{B_{\perp}}{M_{{\rm s}}}\gamma\partial_{z}u_{x}=0.\label{eq:LL_lin_My}
\end{alignat}
\end{widetext}
In the ferromagnet, the stiffness coefficients, mass density, and elastic damping are denoted with tilded quantities to distinguish them from those in the substrate. Equations~(\ref{eq:LL_lin_Mx}, \ref{eq:LL_lin_My}) for the magnetization describe the ferromagnetic resonance in the presence of Gilbert damping and the coupling given by the derivatives of the displacement fields. This is the so-called phonon pumping term. We now proceed to analyze the effect of lattice misalignment in the phonon pumping dynamics.

\section{Birefringence: velocity splitting of the transverse elastic modes}
\label{sec:Velocity-split}

In general, due to the rotation of the elastic strain tensor, the degeneracy of the transverse elastic waves is split giving rise to different velocities for the two modes. This is characterized by the group velocity, which can be found from the dispersion equation. In the investigated structure, the rotation of the strain tensor in the non-magnetic layer has a dominant effect, while the rotation of the strain tensor in the magnetic layer is negligible. The details of the derivation are given in the Appendix~\ref{sec:appendix-Velocity-split}. 

\begin{figure}
\begin{centering}
\includegraphics{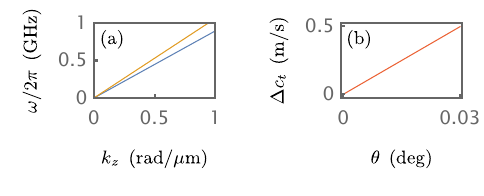} 
\par\end{centering}
\caption{(a) Dispersion relations of elastic waves in an hexagonal crystal for $x$- (blue line) and $y$- (orange line) polarized elastic waves at a relatively large angle of deflection of the strain tensor $\theta=\pi/6$ for visualization purposes. (b) Difference in group velocities of the waves. The calculation parameters correspond to Al$_2$O$_3$ films, and are as follows: $\rho=3970\;{\rm kg/m^{3}}$, $c_{1,1}=5.00073\times10^{11}\;{\rm Pa}$, $c_{3,3}=5.02385\times10^{11}\;{\rm Pa}$, $c_{4,4}=1.51017\times10^{11}\;{\rm Pa}$, $c_{1,2}=1.61672\times10^{11}\;{\rm Pa}$, $c_{1,3}=1.11368\times10^{11}\;{\rm Pa}$, $c_{1,4}=-2.32604\times10^{10}\;{\rm Pa}$, and $\eta=0$. These stiffness are extracted from Ref. \citep{Tefft1966} }
\label{fig:disper_dvg} 
\end{figure}

We obtain three wave numbers corresponding to three linearly polarized modes along $x$, $y$, and $z$, denoted by the corresponding superscript in the following. The wave numbers in the cubic ferromagnetic layer are labeled with the tilde symbol, while the wave numbers without the tilde symbol refer to the hexagonal elastic layer. We find that the exact dispersion relations
\begin{alignat}{1}
k_{z}^{x} & =\sqrt{\mathit{\bar{b}}/3-t_{2}},\label{eq:kxz}\\
k_{z}^{y} & =\sqrt{\mathit{\bar{b}}/3-t_{1}},\label{eq:kyz}\\
k_{z}^{z} & =\sqrt{\mathit{\bar{b}}/3-t_{0}}\label{eq:kzz}
\end{alignat}
undergo significant simplification when $\theta=0$, 
\begin{alignat}{1}
k_{z}^{x} & =\sqrt{\frac{\rho\omega^{2}-{\rm i}\eta\omega}{c_{4,4}}},\label{eq:kxz_th0}\\
k_{z}^{y} & =\sqrt{\frac{\rho\omega^{2}-{\rm i}\eta\omega}{c_{4,4}}},\\
k_{z}^{z} & =\sqrt{\frac{\rho\omega^{2}-{\rm i}\eta\omega}{c_{3,3}}}\,,\label{eq:kzz_th0}
\end{alignat}
that is, when there is no misalignment of the lattices. The expressions for $t_{i}$ ($ i=1,2,3$) and $\mathit{\bar{b}}$ are lengthy and are given in Appendix~\ref{sec:appendix-Velocity-split}. The exact dispersion relations Eqs.~(\ref{eq:kxz}, \ref{eq:kyz}) are plotted in Fig.~\ref{fig:disper_dvg}(a) for a fixed misalignment angle $\theta$. The curves of the $x$- and $y$-polarized elastic waves have distinct slopes indicating birefringence. The dispersion curves are almost linear across a broad range of experimentally relevant parameters (frequency, deflection angle, etc.). We find that the second derivative $\partial^{2}k/\partial\omega^{2}$ is smaller than $10^{-24}\;\text{s}^{2}/\left(\text{rad}\cdot\text{m}\right)$ in the ranges $0<\omega<1\;\text{THz}$, $0<\theta<\pi/4$ for the parameters listed in Fig.~\ref{fig:disper_dvg}.

Since the second derivative $\partial^{2}k/\partial\omega^{2}$ is small, we can approximate Eqs.~(\ref{eq:kxz}-\ref{eq:kzz}) in this linear-dispersion limit as 
\begin{alignat}{1}
k_{{\rm app,}z}^{x} & =\sqrt{\dfrac{\omega^{2}}{c_{tx}^{2}}-{\rm i}\dfrac{\eta\omega}{\rho c_{tx}^{2}}},\\
k_{{\rm app,}z}^{y} & =\sqrt{\dfrac{\omega^{2}}{c_{ty}^{2}}-{\rm i}\dfrac{\eta\omega}{\rho c_{ty}^{2}}},\\
k_{{\rm app},z}^{z} & =\sqrt{\dfrac{\omega^{2}}{c_{l}^{2}}-{\rm i}\dfrac{\eta\omega}{\rho c_{l}^{2}}},
\end{alignat}
where $c_{tx}$, $c_{ty}$ denote the velocities of $x$- and $y$-polarized transverse modes, whereas $c_{l}$ is the velocity of the longitudinal mode. These velocities  depend on all the parameters including the deflection angle $\theta$. Given that the second derivative $\partial^{2}k/\partial\omega^{2}$ is negligibly small, the group and phase velocities are nearly identical, i.e. 
\begin{equation}
c_{tx}=\left(\dfrac{\partial k_{z}^{x}}{\partial\omega}\right)^{-1}\approx\left(\dfrac{k_{z}^{x}}{\omega}\right)^{-1}.
\end{equation}

Considering the specific materials relevant for the experimental setup of Ref.~\citep{Muller2024Chiral}, we note that the splitting $\Delta c_{t}=c_{tx}-c_{ty}$ in the hexagonal elastic layer is greater than the same difference $\Delta\tilde{c}_{t}$, but in the cubic layer of the bilayer. The difference in the hexagonal Al$_2$O$_3$ substrate depending on the deflection angle $\theta$ is plotted in Fig.~\ref{fig:disper_dvg}(b), and it is $\Delta c_{t}\approx0.5\;\text{m}/\text{s}$ at $\theta=0.03^{\circ}$. On the other hand, in the cubic CoFe alloy $\Delta\tilde{c}_{t}\approx-0.001\;\text{m}/\text{s}$ at $\theta=0.03^{\circ}$ for $\tilde{\rho}=8110\;{\rm kg/m^{3}}$, $\tilde{c}_{4,4}=8.15\times10^{10}\;{\rm Pa}.$ The absolute value of the velocities are $c_{tx}=6167.87\;{\rm m/s},$ $c_{ty}=6167.37\;{\rm m/s},$ $\tilde{c}_{tx}\approx\tilde{c}_{ty}=3170\;{\rm m/s}$ at $\theta=0.03^{\circ}$. 

Based on the comparison of hexagonal and cubic bilayers (see Appendix~\ref{sec:hex-cubi}), we conclude that the rotation of the strain tensor stronger affects the velocities of the elastic waves in crystals with lower symmetries and low damping. The investigated structure supports the propagation of linearly polarized elastic waves with different velocities. It leads to lifting of degeneracy and doubling of elastic absorption lines (see sec.~\ref{sec:Num_sim}). Thus, phononic birefringence can be described by the rotation of the elastic strain tensor, and it is more pronounced in crystals with lower symmetries.

Having derived the dispersion equations describing the displacement field, we proceed now to include the boundary conditions, in order to study the influence of the displacement fields on the magnetization dynamics.

\section{Power absorption}

\label{sec:Power-abs}

\subsection{Boundary conditions}

Many experiments exploring the bilayer arrangment depicted in Fig.~\ref{fig:str_rotation}(b) rely on ferromagnetic resonance spectrostopy to investiagte phonon pumping ~\citep{Muller2024Chiral,MullerTemperature2024,an_coherent_2020,an_bright_2022,schlitz_magnetization_2022}. Our goal is therefore to derive the susceptibility as the relation between the exciting magnetic field and magnetization response. The susceptibility can be found with Eqs.~(\ref{eq:LL_lin_Mx}) and~(\ref{eq:LL_lin_My}), but these equations include the derivative of the displacement. Following Ref.~\citep{Streib2018}, we assume that both the applied magnetic field and magnetization are uniform, therefore the response is sensitive only to the spatial average of the fields. Consequently, we can average the derivative over the film thickness 
\begin{alignat}{1}
\partial_{z}u_{x} & =\frac{u_{x}(0)-u_{x}(-d)}{d},\label{eq:dux_via_diff}\\
\partial_{z}u_{y} & =\frac{u_{y}(0)-u_{y}(-d)}{d},\label{eq:duy_via_diff}
\end{alignat}
where $u_{x}(0),$ $u_{x}(-d)$, etc. are the displacements at the boundaries of the ferromagnetic layer.

Before we delve into the details of the boundary problem, let us make a brief remark. In this study, we start from the vector fields of magnetization and displacement. However, we can change the perspective to facilitate a deeper understanding. On one hand, the magnetization is associated with  angular momentum; on the other hand, the displacement fields can carry both linear and angular momentum, as well as relating to the global movement of the structure~\citep{sato_dynamic_2021}. Here we consider only the angular momentum ${\mathbf L}$ arising from the displacement ${\mathbf u}$~\citep{Lifa2014, Garanin2015, Jotaro2018, Ruckriegel2020}
\begin{equation}
\mathbf{L}=\int d^3r\,\rho\mathbf{u}\times\partial_{t}\mathbf{u}.
\end{equation}
Thus, the magnetization excitation dissipates not only due to the Gilbert damping, but also through the current of angular momentum across the interface, leading to the emission of elastic waves, i.e. phonon pumping. Note that this is an interface phenomenon, and it should be distinguished from the bulk magnetoelastic coupling.

We assume that the surfaces of the structure are free, implying that the normal components of the stress tensor are zero. At the $A$ boundary, {[}Fig.~\ref{fig:str_rotation}(a){]}, we have only elastic stress, which vanishes 
\begin{equation}
\sigma_{\nu,z}^{{\rm el}}\left(L\right)=0,\label{eq:boundary_s_A}
\end{equation}
where $\nu=x,y,z.$ At the $B$ boundary, the total stress is the sum of elastic and magnetoelastic contributions 
\begin{equation}
\sigma_{\nu,z}^{{\rm el}}\left(-d\right)+\sigma_{\nu,z}^{{\rm mel}}\left(-d\right)=0,\label{eq:boundary_s_B}
\end{equation}
At the $C$ boundary, the stress tensors and displacements of two layers should match each other 
\begin{alignat}{1}
\sigma_{\nu,z}^{{\rm el}}\left(0^{-}\right)+\sigma_{\nu,z}^{{\rm mel}}\left(0^{-}\right) & =\sigma_{\nu,z}^{{\rm el}}\left(0^{+}\right),\nonumber \\
u_{\nu}\left(0^{-}\right) & =u_{\nu}\left(0^{+}\right),\label{eq:boundary_s_C}
\end{alignat}
We use a standing wave ansatz for the elastic and magnetoelastic layers
\begin{align}
u_{\nu}^{{\rm el}} & =A_{\nu}e^{{\rm i}k_{z}^{\nu}z}+B_{\nu}e^{-{\rm i}k_{z}^{\nu}z},\\
u_{\nu}^{{\rm mel}} & =\tilde{A}_{\nu}e^{{\rm i}\tilde{k}_{z}^{\nu}z}+\tilde{B}_{\nu}e^{-{\rm i}\tilde{k}_{z}^{\nu}z}.
\end{align}
The number of Eqs.~(\ref{eq:boundary_s_A}-\ref{eq:boundary_s_C}) matches the number of unknown amplitudes $A_{\nu},$ $B_{\nu}.$ etc. However, the boundary conditions includes $\sigma_{\nu,z}^{{\rm mel}}$, which is independent of the amplitudes. This term reflects the phonon pumping. Consequently, we can derive a closed expression for the derivatives in Eqs.~(\ref{eq:dux_via_diff}, \ref{eq:duy_via_diff}) 
\begin{align}
\partial_{z}u_{x} & =R_{x}M_{x},\label{eq:du_via_M}\\
\partial_{z}u_{y} & =R_{y}M_{y},
\end{align}
where
\begin{widetext}
\begin{equation}
R_{\nu}=-\frac{B_{\perp}\left[c_{4,4}k_{z}^{\nu}\sin\left(k_{z}^{\nu}L\right)\sin\left(\tilde{k}_{z}^{\nu}d\right)+4\tilde{c}_{4,4}\tilde{k}_{z}^{\nu}\cos\left(k_{z}^{\nu}L\right)\sin^{2}\left(\tilde{k}_{z}^{\nu}d/2\right)\right]}{dM_{{\rm s}}\tilde{c}_{4,4}\tilde{k}_{z}^{\nu}\left[c_{4,4}k_{z}^{\nu}\sin\left(k_{z}^{\nu}L\right)\cos\left(\tilde{k}_{z}^{\nu}d\right)+\tilde{c}_{4,4}\tilde{k}_{z}^{\nu}\cos\left(k_{z}^{\nu}L\right)\sin\left(\tilde{k}_{z}^{\nu}d\right)\right]}.\label{eq:R_u_der}
\end{equation}
\end{widetext}

\subsection{Power absorption}
Ferromagnetic resonance experiments can provide insights into the magnon-phonon coupling and magnetoelastic dynamics via the measurement of the microwave power absorption. This absorption is proportional to the imaginary part of Polder susceptibility $P_{\mathrm{abs}}\propto{\rm Im}\chi$  (Ref.~\citep{sato_dynamic_2021, gurevich_magnetization_2020}), which is the relation between excitation and response of the structure given by $\left(M_{x},M_{y}\right)^{T}=\hat{\chi}\left(H_{x},H_{y}\right)^{T}$ or 
\begin{equation}
\left(\begin{array}{c}
M_{x}\\
M_{y}
\end{array}\right)=\left(\begin{array}{cc}
\chi_{xx} & {\rm i}\chi_{xy}\\
-{\rm i}\chi_{yx} & \chi_{yy}
\end{array}\right)\left(\begin{array}{c}
H_{x}\\
H_{y}
\end{array}\right).\label{eq:sus_mat}
\end{equation}
For $\theta=0$ and $N_{xx}=N_{yy}$ we have $\chi_{xy}=\chi_{yx}$ and $\chi_{xx}=\chi_{yy}$. The latter equality however does not hold if $\theta\neq0,$ due to the non-degeneracy of the transverse linear polarized elastic waves. From Eqs.~(\ref{eq:LL_lin_Mx}, \ref{eq:LL_lin_My}) we can find the susceptibility tensor explicitly
\begin{equation}
\hat{\chi}=\gamma\mu_{0}M_{{\rm s}}U\left(\begin{array}{cc}
T_{zx}\left(1+\gamma B_{\perp}R_{y}T_{zy}\right) & i\omega T_{zx}T_{zy}\\
-i\omega T_{zx}T_{zy} & T_{zy}\left(1+\gamma B_{\perp}R_{x}T_{zx}\right)
\end{array}\right),\label{eq:sus}
\end{equation}
where 

\begin{eqnarray}
U & = & \left[1+\gamma B_{\perp}R_{x}T_{zx}\left(\gamma B_{\perp}R_{y}T_{zy}+1\right)\right.\\
 &  & \left.+\gamma B_{\perp}R_{y}T_{zy}-\omega^{2}T_{zx}T_{zy}\right]^{-1},\\
T_{zx} & = & \left(i\alpha\omega+\gamma\mu_{0}H_{0}+\frac{2\gamma K_{1}}{M_{{\rm s}}}\right.\\
 &  & \left.+\gamma\mu_{0}M_{{\rm s}}N_{xx}-\gamma\mu_{0}M_{{\rm s}}N_{zz}\right)^{-1},\\
T_{zy} & = & \left(i\alpha\omega+\gamma\mu_{0}H_{0}+\frac{2\gamma K_{1}}{M_{{\rm s}}}\right.\\
 &  & \left.+\gamma\mu_{0}M_{{\rm s}}N_{yy}-\gamma\mu_{0}M_{{\rm s}}N_{zz}\right)^{-1}.\label{eq:sus_U}
\end{eqnarray}

In the investigated structure, the ferromagnetic mode is coupled to the elastic mode, resulting in a complex dependence of the susceptibility on the magnetic field, deviating from the standard Kittel equations.

\subsection{Numerical simulation of absorption}

\label{sec:Num_sim}

\begin{figure}
\begin{centering}
\includegraphics{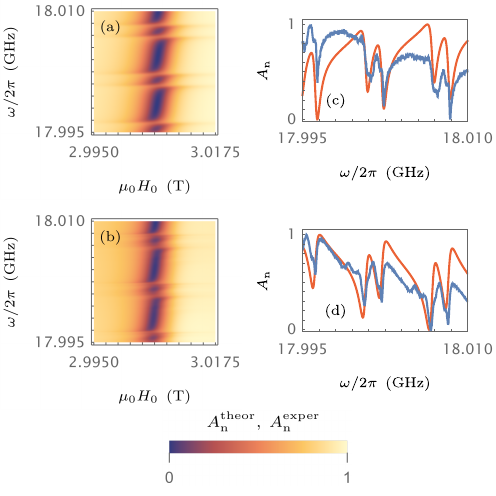} 
\par\end{centering}
\caption{(a) Normalized theoretical power absorption $A_{\rm n}^{\text{theor}}$. (b) Normalized microwave transmission magnitude $A_{\rm n}^{\text{exper}}$ related to the absorption $\left|S_{21}\right|$ from Ref.~\citep{Muller2024Chiral}. (c) and (d) Comparison between the experimental data (blue) and theoretical calculation (red) of the relative absorption at $\mu_{0}H_{0}=3\;{\rm T}$ (c) and $\mu_{0}H_{0}=3.01\;{\rm T}$ (d). The experimental and theoretical data are normalized. The calculation parameters correspond to the  Al$_2$O$_3$ and Co$_{25}$Fe$_{75}$ layers and they are $\gamma/\left(2\pi\right)=29.80\;{\rm GHz/T}$, $K_{1}=0$, $M_{{\rm s}}=1.91083\times10^{6}\;{\rm A/m},$ $L=510\;\mu{\rm m}$, $d=30\;{\rm nm}$, $\rho=3970\;{\rm kg/m^{3}}$, $\tilde{\rho}=8110\;{\rm kg/m^{3}}$, $c_{1,1}=5.00073\times10^{11}\;{\rm Pa}$, $c_{3,3}=5.02385\times10^{11}\;{\rm Pa}$, $c_{4,4}=1.51017\times10^{11}\;{\rm Pa}$, $c_{1,2}=1.61672\times10^{11}\;{\rm Pa}$, $c_{1,3}=1.11368\times10^{11}\;{\rm Pa}$, $c_{1,4}=-2.32604\times10^{10}\;{\rm Pa}$, $\tilde{c}_{4,4}=8.15\times10^{10}\;{\rm Pa}$, $c_{tx}=6167.87\;{\rm m/s},$ $c_{ty}=6167.37\;{\rm m/s},$ $\tilde{c}_{tx}=\tilde{c}_{ty}=3170\;{\rm m/s}$ $B_{\perp}=15.7\times10^{6}\;{\rm J/m^{3}}$, $\alpha=0.004$, $\eta=2\pi\times2.03\;{\rm GN/m^{4}}$, $\phi_e = 0.22798\;{\rm rad}$, $N_{xx}=N_{yy}=0$, $N_{zz}=1$ and $\theta=0.03^{\circ}$.}
\label{fig:absorbtion} 
\end{figure}

The relative absorption given by $A={\rm Im}\left(\chi_{xx}+\chi_{yy}\right)$ (Ref.~\citep{sato_dynamic_2021}) is plotted in Fig.~\ref{fig:absorbtion}(a). The ferromagnetic mode is represented by the broad, nearly vertical dark blue absorption line. The almost vertical slope of this mode is primarily a result of the scale shown in the figure, since the slope of the bare magnetic mode is comparatively large (given by $\gamma/\left(2\pi\right)=29.80\;{\rm GHz/T}$). The repeating pairs of horizontal features are elastic modes. We can identify two effects: (i) birefringence of elastic modes, (ii) hybridization of elastic and magnetic modes.

Let us address the birefringence first. The propagating elastic waves reflect from boundaries at $-d$ and $L$ (see Fig.~\ref{fig:str_rotation}), and this reflection results in the formation of standing elastic waves. Therefore the wave vector is discretized as $n\pi/\left(L+d\right)$ with $n$ being an integer. We observe this spacing between the pairs of elastic modes. For every $n$, there are two transverse linearly polarized elastic standing waves, each with a distinct velocity (or wavelength). This arises from the relatively small deflection angle $\theta$, which lifts the degeneracy of these elastic waves. Consequently, the two modes are distinguishable by their polarization, demonstrating birefringence.

The coupling between the ferromagnetic and elastic modes manifest as modification of the Polder susceptibility or the microwave absorption. Fig.~\ref{fig:absorbtion} shows the pronounced case with avoided crossings near resonance of the elastic and magnetic modes, which suggest that the magnetoelastic coupling can result in a mode hybridization. The resulting magnon-phonon polariton presents dispersion, that is, the elastic waves acquire magnetic properties, making their frequencies dependent on the magnetic field. Note that the observed hybridization differs from phase synchronization, which requires phase matching of two waves, meaning the frequencies \emph{and} wave vectors must coincide. The observed hybridization in this case occurs due to the boundary conditions, which describe the exchange of angular momentum between elastic and magnetic subsystems at the interface.

As a next step, we compare our results to the microwave transmission experiments conducted in Ref.~\citep{Muller2024Chiral} {[}see Fig.~\ref{fig:absorbtion}(b){]}. 
%In the  microwave transmission experiments, the structure is coupled inductively to a coplanar waveguide and a static external magnetic field is applied. 
The structure consists of a \emph{c}-plane sapphire substrate with $L=510\;\mu{\rm m}$ and finite miscut angle $\theta$, onto which a thin film layer of Co$_{25}$Fe$_{75}$ ($d=30\;{\rm nm}$) has been deposited. The experimental data has been obtained at a temperature of $5\;{\rm K}$ and a center frequency of the microwave of $18\;{\rm GHz}$ in a static external magnetic field perpendicular to the plane of the sample. Further details on the experimental measurement approach can be found in Ref.~\citep{Muller2024Chiral}.

%To enable a direct comparison between simulation and experiment, the transmission data were normalized to values between 0 and 1. We also introduced an additional phase shift  $\phi_e$ following the approach in Ref.~\citep{Maier-Flaig}, thus the normalized theoretical power absorption is proportional to

%\begin{equation}
%	P=Z^{-1}{\rm Im}[ e^{i\phi_e}( \chi_{xx}+\chi_{yy})],\label{eq:power_shift} 
%\end{equation}
%where $Z$ represents the maximum power absorption.
%Additionally, we provide the comparison between the experimental and theoretical data at specific values of the magnetic field. {[}Fig.~\ref{fig:absorbtion}(c,d){]}. The quantitative fit is observed, indicating that the proposed theory accurately captures the magnetoelastic dynamics in the structure.

To enable a direct comparison between  simulation and experiment, we first need to review how the measured complex microwave transmission $S_{21}$ links to the dynamic magnetic susceptibility $\chi$. In the experiment, the magnetic element couples inductively to the microwave delivery and readout circuit, in form of a coplanar waveguide. The complex $S_{21}$ is given by~\citep{Schoen2015}
\begin{equation}
    S_{21}(H_0)_{\omega}=A + B H_0 + \sum_{n}^N C e^{i\phi} \chi_n(\omega, H_0),
\end{equation}
where $A$ and $B$ describe the background of the microwave transmission up to linear order in the magnetic field $H_0$, while $C$ parametrizes the coupling of the magnetic element to the microwave circuit \footnote{We disregard the linear frequency dependence of this coupling since we are fitting in a small frequency range compared to the center frequency $f_0$}. In addition, the phase $\phi$ accounts for phase shifts which originate from the electric length. 
Fig.~\ref{fig:absorbtion}(b) shows the normalized magnitude of $S_{21}$, corroborating the simulation presented in panel (a). The normalization is done by taking the leftmost vertical slice at constant magnetic field of the measured spectrum of the transmission parameter $S_{21}$ and divide the whole spectrum by this slice. A quantitative comparison is presented in the form of fixed magnetic field cuts in the panels (c) and (d). To this end, we convert the complex susceptibility [see Eq.~(\ref{eq:sus})] to normalized magnitude spectrum of $S_{21}$, demonstrating that the proposed theory accurately captures the magnetoelastic dynamics in the structure. 

\section{Polarization}

\label{sec:Polarization}

Studying the polarization of the hybrid modes in the structure provides information on the angular momentum transfer between the elastic and magnetic modes. The polarization can be characterized by the Stokes parameters, a standard tool in optics to e.g. analyze the angular momentum of light in waveguides, optical fibers, etc. \citep{bliokh2015transverse, Todd2016}. For the magnetization, we can introduce the normalized Stokes parameter $V$ following \citep{McMaster1961} as:
\begin{equation}
\begin{aligned}V & =-2\frac{{\rm Im}\left(M_{x}M_{y}^{*}\right)}{\left|M_{x}\right|^{2}+\left|M_{y}\right|^{2}}\end{aligned}
\label{eq:Stokes_V}
\end{equation}
This parameter equals $-1$ for a right-handed circular polarized field and $+1$ for a left-handed polarized field. The dependence of polarization on the exciting magnetic field can be found by substituting the magnetization with its expression given by the susceptibility, Eq.~\eqref{eq:sus_mat}, into Eq.~\eqref{eq:Stokes_V}. 

\begin{figure}
\begin{centering}
\includegraphics[width=8.3cm]{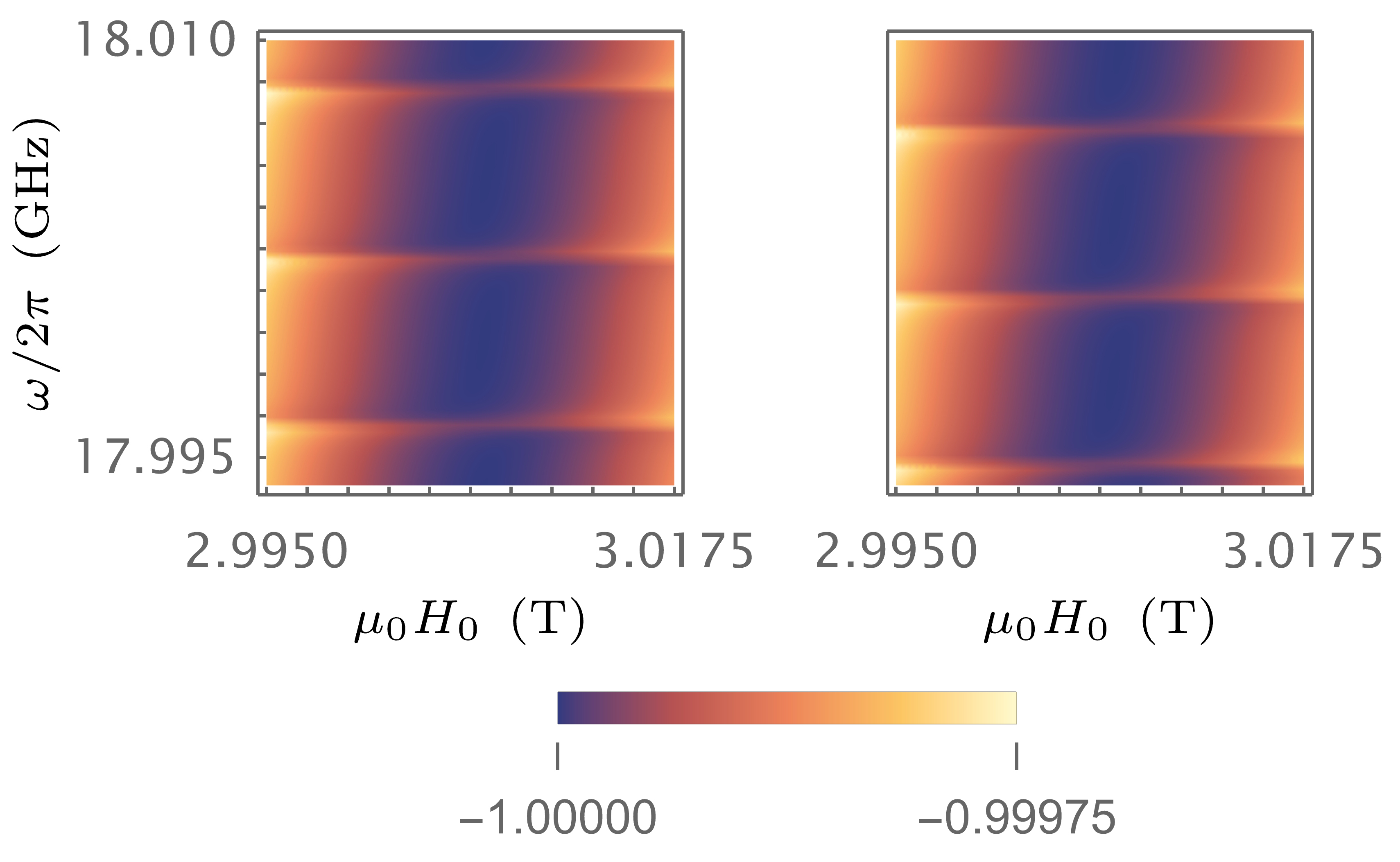} 
\par\end{centering}
\caption{Stokes parameter $V$ of magnetization at different polarization of exciting magnetic field: (a) $\emph{x}$-polarized field (b) $\emph{y}$-polarized field. The calculation parameters are the same as for Fig.$~$$\ref{fig:absorbtion}$.}
\label{fig:Stokes} 
\end{figure}

We analyze the polarization of the response for an $x$- and $y$-linear polarized exciting magnetic field separately. The Fig.~\ref{fig:Stokes}(a) illustrates how the circular polarization of the magnetization is affected by the $x$-polarized magnetic field, while the Fig.~\ref{fig:Stokes}(b) demonstrates the effect of the $y$-polarized magnetic field. For the calculation of the circular Stokes parameter $V$, only the orthogonality between the $x$ and $y$ polarization axes is relevant, not their orientation in space. The frequency of the periodic bright and approximately horizontal lines is shifted between the two panels, i.e. the frequency of the $x$-elastic waves is greater than the frequency of $y$-elastic waves. We conclude that the double anti-crossings in Fig.~\eqref{fig:absorbtion}(a) result from the velocity difference of $x$- and $y$-linear polarized elastic waves excited by different components of the magnetic field. Note that the velocity difference is the manifestation of the birefringence.

By comparing Figs.~\eqref{fig:absorbtion}(a) with \ref{fig:Stokes}(a) and (b), we observe an additional effect of this coupling: the hybrid elastic-magnetic mode acquires linear polarization in the anti-crossing region. This linear polarization is reflected in the magnetic response (as characterized by Eq.~\eqref{eq:Stokes_V} ) and it is more pronounced as the excitation frequency detunes further from the eigenfrequency, with the mode becoming more elastic than magnetic. In this regime, the response shows weak linear polarization, arising from the coupling to linearly polarized elastic shear waves.

\section{Conclusion}
\label{sec:conclusions} 

In this study, we have presented a theoretical framework that captures both phonon pumping and the birefringence arising from misaligned crystal and magnetic symmetry axes in a bilayer structure. Our model shows quantitative agreement with the experimental findings of Ref.~\cite{Muller2024Chiral}, demonstrating that phononic birefringence is manifested in the splitting of the two transverse elastic modes and that the ensuing magnon-phonon hybridization imparts the otherwise nonmagnetic phonon modes with a magnetic field dependence. Conversely, the circularly polarized Kittel mode acquires a finite linear polarization from the phonons, illustrating the bidirectional nature of magnon-phonon coupling in these structures.

Beyond confirming that the misalignment of elastic and magnetic symmetries lifts the degeneracy of transverse shear modes, our results highlight that boundary conditions and phonon pumping have a strong impact on the mode dispersion, even in high-symmetry crystals. These effects are important for understanding the angular momentum transport and for engineering coherent magnon-phonon interactions over long distances.

Future studies may extend the present theory to a wider variety of crystal symmetries, exploit mode- and polarization-selective control of the hybrid states, and investigate the impact of additional forms of anisotropy or strain-mediated coupling in layered heterostructures. From a technological perspective, these insights can be leveraged in spintronics, including magnon transport, and in quantum acoustic transduction, where precise control of phonon polarization and propagation is essential for device performance.

\section{Acknowledgments}
We acknowledge financial support by the German Federal Ministry of Education and Research (BMBF) project QECHQS (Grant No. 16KIS1590K). We acknowledge funding from the Deutsche Forschungsgemeinschaft (DFG, German Research Foundation) under Germany’s Excellence Strategy—EXC-2111-390814868 and the Transregio “Constrained Quantum Matter” (TRR 360, Project-ID 492547816). This research is part of the Munich Quantum Valley, which is supported by the Bavarian State Government with funds from the Hightech Agenda Bayern Plus.

\section{Data availability}
The data that support the findings of this study are available from the corresponding author upon reasonable request.

\appendix
%dummy comment inserted by tex2lyx to ensure that this paragraph is not empty%dummy comment inserted by tex2lyx to ensure that this paragraph is not empty

\section{Details of velocity splitting calculations}

\label{sec:appendix-Velocity-split}

Here we provide the details of the calculation of the velocity splitting. We start from the derivation of the dispersion Eq. from the Hooke's Eqs.~(\ref{eq:Hooke_LL_el1}-\ref{eq:Hooke_LL_el3}). Note that the following calculations are also applicable to Eqs.~(\ref{eq:Hooke_LL_mel1}-\ref{eq:Hooke_LL_mel3}) using the transformation given by Eqs.~(\ref{eq:fromHexToCubic}).

The system of Eqs.~(\ref{eq:Hooke_LL_el1}-\ref{eq:Hooke_LL_el3}) can be written as 
\begin{alignat}{1}
 & \left(\rho\omega^{2}-{\rm i}\eta\omega\right)u_{x}+\mathbb{D}_{\text{x}}^{\text{ux}}\partial_{zz}u_{x}+\mathbb{D}_{\text{y}}^{\text{ux}}\partial_{zz}u_{y}+\mathbb{D}_{\text{z}}^{\text{ux}}\partial_{zz}u_{z}=0,\label{eq:Hooke_LL_u_D1}\\
 & \left(\rho\omega^{2}-{\rm i}\eta\omega\right)u_{y}+\mathbb{D}_{\text{x}}^{\text{uy}}\partial_{zz}u_{x}+\mathbb{D}_{\text{y}}^{\text{uy}}\partial_{zz}u_{y}+\mathbb{D}_{\text{z}}^{\text{uy}}\partial_{zz}u_{z}=0,\\
 & \left(\rho\omega^{2}-{\rm i}\eta\omega\right)u_{z}+\mathbb{D}_{\text{x}}^{\text{uz}}\partial_{zz}u_{x}+\mathbb{D}_{\text{y}}^{\text{uz}}\partial_{zz}u_{y}+\mathbb{D}_{\text{z}}^{\text{uz}}\partial_{zz}u_{z}=0,\label{eq:Hooke_LL_u_D3}
\end{alignat}
where

\begin{alignat}{1}
\mathbb{D}_{\text{x}}^{\text{ux}} & =c_{4,4}\cos\left(2\theta\right),\\
\mathbb{D}_{\text{y}}^{\text{ux}} & =-c_{1,4}\sin\left(\theta\right),\\
\mathbb{D}_{\text{z}}^{\text{ux}} & =-c_{4,4}\sin\left(2\theta\right),\\
\mathbb{D}_{\text{x}}^{\text{uy}} & =-c_{1,4}\cos\left(\theta\right)\sin\left(\theta\right),\\
\mathbb{D}_{\text{y}}^{\text{uy}} & =c_{4,4}\cos\left(\theta\right),\\
\mathbb{D}_{\text{z}}^{\text{uy}} & =c_{1,4}\sin^{2}\left(\theta\right),\\
\mathbb{D}_{\text{x}}^{\text{uz}} & =\left(c_{3,3}-c_{1,3}\right)\cos\left(\theta\right)\sin\left(\theta\right),\\
\mathbb{D}_{\text{y}}^{\text{uz}} & =0,\\
\mathbb{D}_{\text{z}}^{\text{uz}} & =c_{1,3}\sin^{2}\left(\theta\right)+c_{3,3}\cos^{2}\left(\theta\right).
\end{alignat}
Note that Eqs.~(\ref{eq:Hooke_LL_u_D1}-\ref{eq:Hooke_LL_u_D3}) at $\theta=0$ are decoupled 
\begin{alignat}{1}
\partial_{zz}u_{x} & =\frac{-\rho\omega^{2}+{\rm i}\eta\omega}{c_{4,4}}u_{x},\label{eq:Hooke_LL_u_th0_1}\\
\partial_{zz}u_{y} & =\frac{-\rho\omega^{2}+{\rm i}\eta\omega}{c_{4,4}}u_{y},\\
\partial_{zz}u_{z} & =\frac{-\rho\omega^{2}+{\rm i}\eta\omega}{c_{3,3}}u_{z},\label{eq:Hooke_LL_u_th0_3}
\end{alignat}
This is not the case if the elastic tensor is rotated, i.e. the symmetry of elastic and magnetic properties are not perfectly aligned. Writing Eqs.~(\ref{eq:Hooke_LL_u_D1}-\ref{eq:Hooke_LL_u_D3}) in the form 
\begin{alignat}{1}
\partial_{zz}u_{x} & =\dfrac{\left|\begin{array}{ccc}
\mathbb{B}^{{\rm ux}} & \mathbb{D}_{\text{y}}^{\text{ux}} & \mathbb{D}_{\text{z}}^{\text{ux}}\\
\mathbb{B}^{{\rm uy}} & \mathbb{D}_{\text{y}}^{\text{uy}} & \mathbb{D}_{\text{z}}^{\text{uy}}\\
\mathbb{B}^{{\rm uz}} & \mathbb{D}_{\text{y}}^{\text{uz}} & \mathbb{D}_{\text{z}}^{\text{uz}}
\end{array}\right|}{\mathbb{\tilde{D}}},\label{eq:Hooke_LL_u_Ddet1}\\
\partial_{zz}u_{y} & =\dfrac{\left|\begin{array}{ccc}
\mathbb{D}_{\text{x}}^{\text{ux}} & \mathbb{B}^{{\rm ux}} & \mathbb{D}_{\text{z}}^{\text{ux}}\\
\mathbb{D}_{\text{x}}^{\text{uy}} & \mathbb{B}^{{\rm uy}} & \mathbb{D}_{\text{z}}^{\text{uy}}\\
\mathbb{D}_{\text{x}}^{\text{uz}} & \mathbb{B}^{{\rm uz}} & \mathbb{D}_{\text{z}}^{\text{uz}}
\end{array}\right|}{\mathbb{\tilde{D}}},\\
\partial_{zz}u_{z} & =\dfrac{\left|\begin{array}{ccc}
\mathbb{D}_{\text{x}}^{\text{ux}} & \mathbb{D}_{\text{y}}^{\text{ux}} & \mathbb{B}^{{\rm ux}}\\
\mathbb{D}_{\text{x}}^{\text{uy}} & \mathbb{D}_{\text{y}}^{\text{uy}} & \mathbb{B}^{{\rm uy}}\\
\mathbb{D}_{\text{x}}^{\text{uz}} & \mathbb{D}_{\text{y}}^{\text{uz}} & \mathbb{B}^{{\rm uz}}
\end{array}\right|}{\mathbb{\tilde{D}}},\label{eq:Hooke_LL_u_Ddet3}
\end{alignat}
where 
\begin{alignat}{1}
\mathbb{\tilde{D}} & =\left|\begin{array}{ccc}
\mathbb{D}_{\text{x}}^{\text{ux}} & \mathbb{D}_{\text{y}}^{\text{ux}} & \mathbb{D}_{\text{z}}^{\text{ux}}\\
\mathbb{D}_{\text{x}}^{\text{uy}} & \mathbb{D}_{\text{y}}^{\text{uy}} & \mathbb{D}_{\text{z}}^{\text{uy}}\\
\mathbb{D}_{\text{x}}^{\text{uz}} & \mathbb{D}_{\text{y}}^{\text{uz}} & \mathbb{D}_{\text{z}}^{\text{uz}}
\end{array}\right|,\\
\mathbb{B}^{{\rm ux}} & =\left(-\rho\omega^{2}+{\rm i}\eta\omega\right)u_{x},\\
\mathbb{B}^{{\rm uy}} & =\left(-\rho\omega^{2}+{\rm i}\eta\omega\right)u_{y},\\
\mathbb{B}^{{\rm uz}} & =\left(-\rho\omega^{2}+{\rm i}\eta\omega\right)u_{z}.
\end{alignat}

Eqs.~(\ref{eq:Hooke_LL_u_Ddet1}-\ref{eq:Hooke_LL_u_Ddet3}) can be written as 
\begin{alignat}{1}
\partial_{zz}u_{x} & =A_{{\rm x}}^{{\rm ux}}u_{x}+A_{{\rm y}}^{{\rm ux}}u_{y}+A_{{\rm z}}^{{\rm ux}}u_{z},\label{eq:Hooke_LL_u_A1}\\
\partial_{zz}u_{y} & =A_{{\rm x}}^{{\rm uy}}u_{x}+A_{{\rm y}}^{{\rm uy}}u_{y}+A_{{\rm z}}^{{\rm uy}}u_{z},\\
\partial_{zz}u_{z} & =A_{{\rm x}}^{{\rm uz}}u_{x}+A_{{\rm y}}^{{\rm uz}}u_{y}+A_{{\rm z}}^{{\rm uz}}u_{z}.\label{eq:Hooke_LL_u_A3}
\end{alignat}
The explicit form of the $A$ coefficients can be determined using (\ref{eq:Hooke_LL_u_Ddet1}-\ref{eq:Hooke_LL_u_Ddet3}).

We then move on to analyze of the dispersion given by the system of Eqs.~(\ref{eq:Hooke_LL_u_A1}-\ref{eq:Hooke_LL_u_A3}). The second derivative in these equations can be transformed into a first derivative by employing an auxiliary variable. This transformation leads to a doubling of the number of Eqs., resulting a total of 6 Eqs. This allows us to express the dispersion Eq. as follows: 
\begin{equation}
\det\left[\left(\begin{array}{cccccc}
0 & 0 & 0 & 1 & 0 & 0\\
0 & 0 & 0 & 0 & 1 & 0\\
0 & 0 & 0 & 0 & 0 & 1\\
A_{{\rm x}}^{{\rm ux}} & A_{{\rm y}}^{{\rm ux}} & A_{{\rm z}}^{{\rm ux}} & 0 & 0 & 0\\
A_{{\rm x}}^{{\rm uy}} & A_{{\rm y}}^{{\rm uy}} & A_{{\rm z}}^{{\rm uy}} & 0 & 0 & 0\\
A_{{\rm x}}^{{\rm uz}} & A_{{\rm y}}^{{\rm uz}} & A_{{\rm z}}^{{\rm uz}} & 0 & 0 & 0
\end{array}\right)-ik_{z}\mathbb{{1}}\right]=0,
\end{equation}
which in turn yields 
\begin{alignat}{1}
k_{z}^{6}+k_{z}^{4}\bar{b}+k_{z}^{2}\bar{c}+\bar{d}=0,\label{eq:bicubic_kz}
\end{alignat}
where 
\begin{equation}
\begin{aligned}\bar{d}= & A_{{\rm y}}^{{\rm uy}}A_{{\rm x}}^{{\rm uz}}A_{{\rm z}}^{{\rm ux}}-A_{{\rm x}}^{{\rm uy}}A_{{\rm z}}^{{\rm ux}}A_{{\rm y}}^{{\rm uz}}-A_{{\rm y}}^{{\rm ux}}A_{{\rm x}}^{{\rm uz}}A_{{\rm z}}^{{\rm uy}}\\
 & +A_{{\rm x}}^{{\rm ux}}A_{{\rm y}}^{{\rm uz}}A_{{\rm z}}^{{\rm uy}}+A_{{\rm z}}^{{\rm uz}}A_{{\rm x}}^{{\rm uy}}A_{{\rm y}}^{{\rm ux}}-A_{{\rm x}}^{{\rm ux}}A_{{\rm y}}^{{\rm uy}}A_{{\rm z}}^{{\rm uz}},\\
\bar{c}= & -A_{{\rm x}}^{{\rm uy}}A_{{\rm y}}^{{\rm ux}}+A_{{\rm x}}^{{\rm ux}}A_{{\rm y}}^{{\rm uy}}-A_{{\rm x}}^{{\rm uz}}A_{{\rm z}}^{{\rm ux}}\\
 & A_{{\rm x}}^{{\rm ux}}A_{{\rm z}}^{{\rm uz}}-A_{{\rm y}}^{{\rm uz}}A_{{\rm z}}^{{\rm uy}}+A_{{\rm y}}^{{\rm uy}}A_{{\rm z}}^{{\rm uz}},\\
\bar{b}= & -A_{{\rm x}}^{{\rm ux}}-A_{{\rm y}}^{{\rm uy}}-A_{{\rm z}}^{{\rm uz}}.
\end{aligned}
\label{eq:cubic_abcd}
\end{equation}
The bi-cubic Eq.~\eqref{eq:bicubic_kz} has three pairs of roots, and we select the positive ones for further analysis. We define three wave numbers corresponding to $x-$, $y-$, and $z-$linearly polarized waves, denoted by the superscript
\begin{alignat}{1}
k_{z}^{x} & =\sqrt{\mathit{\bar{b}}/3-t_{2}},\\
k_{z}^{y} & =\sqrt{\mathit{\bar{b}}/3-t_{1}},\\
k_{z}^{z} & =\sqrt{\mathit{\bar{b}}/3-t_{0}}
\end{alignat}
where 
\begin{alignat}{1}
t_{j} & =2\sqrt{-\frac{p}{3}}\cos\left[\frac{1}{3}\arccos\left(\frac{3q}{2p}\sqrt{-\frac{3}{p}}\right)-\frac{2\pi}{3}j\right],\\
p & =\mathit{\bar{c}}-\frac{\mathit{\mathit{\bar{b}}}^{2}}{3},\\
q & =\mathit{\bar{d}}+\frac{2\mathit{\mathit{\bar{b}}}^{3}-9\mathit{\mathit{\bar{b}}}\mathit{\bar{c}}}{27}.
\end{alignat}
The same derivation can be performed for the cubic crystal by replacing $\mathbb{D}$ coefficients. We denote the wave numbers in the cubic ferromagnetic layer as $\tilde{k}_{z}^{x},$ $\tilde{k}_{z}^{y},$ and $\tilde{k}_{z}^{z}$, while the same wave number but without the tilde symbol refers to the hexagonal elastic layer. 

\section{Comparison of hexagonal and cubic elastic systems}\label{sec:hex-cubi}

Based on numerical calculations we conclude that the phononic birefringence is more pronounced in elastic crystals with low symmetry. These low-symmetry crystals are described by a stiffness tensor with more non-zero elements. The rotation Eq.~\eqref{eq:epsRotation} introduce trigonometric functions into these elements in subsequent equations making the crystal more "sensitive" to deflection angle $\theta.$ In contrast, high-symmetry crystals are less affected by the rotation, resulting in weaker birefringence. 

\begin{figure}[H]
\begin{centering}
\includegraphics{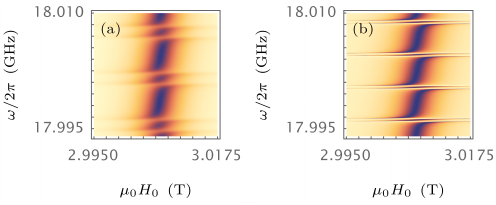}
\par\end{centering}
\caption{Susceptibility of hexagonal (a) and cubic (b) bilayers. The parameters for hexagonal structure consisting of Al$_2$O$_3$ and Co$_{25}$Fe$_{75}$ layers are the same as in Fig.~\ref{fig:absorbtion}. The parameters for cubic structure comprised of  Si and Co$_{25}$Fe$_{75}$ layers were taken as $\gamma/\left(2\pi\right)=29.80\;{\rm GHz/T}$, $K_{1}=0$, $M_{{\rm s}}=1.91083\times10^{6}\;{\rm A/m},$ $L=675\;\mu{\rm m}$, $d=30\;{\rm nm}$, $\rho=2330\;{\rm kg/m^{3}}$, $\tilde{\rho}=8110\;{\rm kg/m^{3}}$, $c_{1,1}=c_{3,3}=161\;{\rm GPa}$, $c_{4,4}=76.1\;{\rm GPa}$, $c_{1,2}=c_{1,3}=64\;{\rm GPa}$, $\tilde{c}_{4,4}=8.15\times10^{10}\;{\rm Pa}$, $B_{\perp}=19.1\times10^{6}\;{\rm J/m^{3}}$, $\alpha=0.004$, $\eta=2.8\times10^{5}\;{\rm N/m^{4}}$, $N_{xx}=N_{yy}=0$, $N_{zz}=1$ and $\theta=0.03^{\circ}$.}
\label{fig:hex-cubic}
\end{figure}

As an example we compare susceptibility of hexagonal and cubic bilayers. The former is the same as displayed in Fig.~\ref{fig:absorbtion}, while the latter consists of elastic cubic Si and magnetoelastic cubic CoFe alloy as used in the experiment \citep{Muller2024Chiral}. The results are shown in Fig.~\ref{fig:hex-cubic}. Two effects are apparent. First, equivalent deflection angles in both bilayers lead to weaker elastic mode splitting in the cubic structure. Second, the visibility of the splitting is strongly influenced by the linewidth of the elastic modes. This highlights that the interplay between structural parameters can significantly enhance the prominence of birefringence.

\bibliographystyle{apsrev4-2}
\bibliography{Ref}

\end{document}